# Imaging polar and dipolar sources of geophysical anomalies by probability tomography. Part II: Application to the Vesuvius volcanic area

**Paolo Mauriello** [1] and **Domenico Patella** [2]

[1] *Department of Science and Technology for Environment and Territory, University of Molise, Campobasso, Italy*
*(E-mail: mauriello@unimol.it)*

[2] *Department of Physical Sciences, University Federico II, Naples, Italy*
*(E-mail: patella@na.infn.it)*

**ABSTRACT**

In the previous part I, we have developed the generalized theory of the probability tomography method to image polar and dipolar sources of a vector or scalar geophysical anomaly field. The purpose of the new method was to improve the core-and-boundary resolution of the most probable buried sources of the anomalies detected in a datum domain. In this paper, which constitutes the part II of the same study, an application of the new approach to the Vesuvius volcano (Naples, Italy) is illustrated in detail by analyzing geoelectrical, self-potential and gravity datasets collected over the whole volcanic area. The purpose is to get new insights into the shallow structure and hydrothermal system of Vesuvius, and the deep geometry of the tectonic depression within which the volcano grew.

## INTRODUCTION

In the first part of this study (Mauriello and Patella, 2006), from now onward indicated with MP-I, we have illustrated the theory of the 3D probability tomography, under the assumption that any geophysical dataset can be interpreted as the response of a double set of hidden sources, schematized as poles and dipoles. Source pole and dipole occurrence probability functions (SPOP and SDOP, respectively) were introduced to obtain a core-and-boundary imaging of the most probable sources of the detected geophysical anomalies.

In this second part, the results from an application to the Mt. Somma-Vesuvius volcano (Naples, Italy) are presented. To avoid repetitions, all previous equations and figures needed to understand this application, will be recalled referring to the numbering used in MP-I.

Geophysics is currently applied in volcanology to define the physical and geometric features of a volcanic apparatus and to study feeding and plumbing systems.

Mt. Somma-Vesuvius is among the most surveyed active volcanoes in the world for the great concern due to the high level of urbanization existing all around its slopes, closest to the city of Naples. A comprehensive collection of the most recent studies about the volcanic structure and dynamics is given by Spera *et al*. (1998).

The geophysical evidences achieved so far indicate that Vesuvius grew inside a depression of a carbonate basement with normal crustal density, filled with less dense sediments. The shallower portion of the volcano has been interpreted as a central plumbing system surrounded by slowly cooled dikes. The summit cone is likely made of altered volcanics. Moreover, a high-to-low velocity and resistivity boundary estimated within a 8-10 km depth range beneath the Vesuvius cone has been assumed as the top of a magma chamber. Though not yet geometrically and structurally well constrained, these features indicate that Vesuvius is still to be considered a highly hazardous volcano.

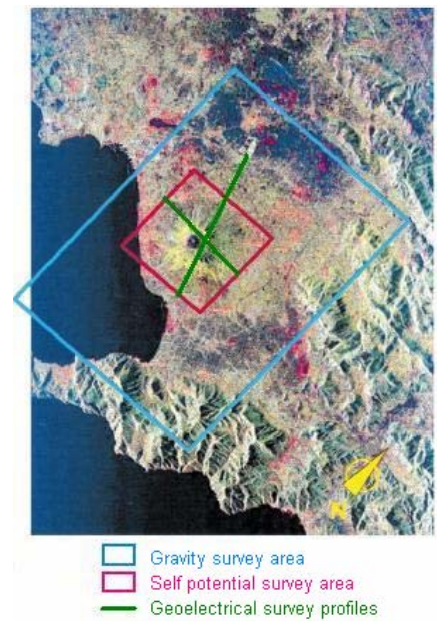

**Figure 1** A satellite map of the Vesuvius volcanic district with localization of the self-potential and gravity survey areas and the geoelectrical profiles.

These results were got using standard modeling and inversion tools, except for the geoelectric, gravity and self-potential data, which were analyzed by the original formulation of the SPOP tomography (Patella, 1997; Patella and Mauriello, 1999; Iuliano *et al.*, 2002).

In the next sections, we give some new results from a joint application of the SPOP and SDOP tomography to the geoelectric, self-potential and gravity datasets on Vesuvius.

A satellite map of the volcanic area and the zones which have been investigated by the methods at issue are reported in figure 1.

## GEOELECTRIC TOMOGRAPHIES

Figure 2 displays the geoelectrical apparent resistivity (GAR) pseudosections across the S-N (upper) and E-W (lower) profiles drawn in figure 1. The GAR data were got using bipoles 500 m long. Each value was assigned at a pseudodepth equal to half the spacing between the centers of the emitting and receiving bipoles, along the median axis across the line jointing the two bipoles. Of course, this standard rule was occasionally adjusted to account for altitude variations along the profiles. The step-by-step displacement of the two bipoles, together with a step-by-step increase of the spacing between the two bipoles provided a dense network of data points in the vertical pseudosection across each profile.

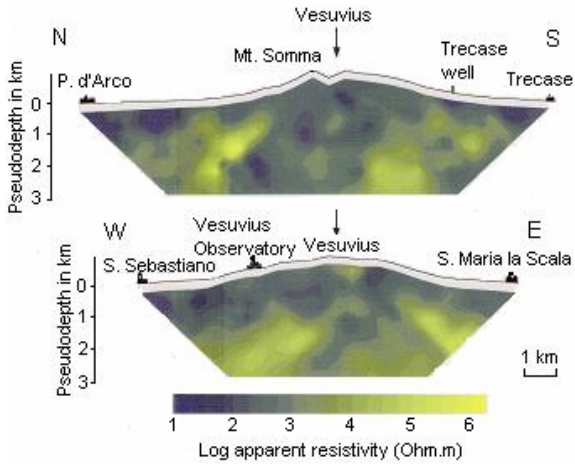

**Figure 2** Geoelectrical apparent resistivity pseudo-sections relative to the N-S (top) and W-E (bottom) profiles shown in figure 1.

Figure 4 displays the resultant geoelectrical SPOP tomographies elaborated from the apparent resistivity pseudosections across the S-N and E-W profiles drawn in figure 1. These SPOP images have been drawn using the approach by Mauriello and Patella (1999), which in the frame of the generalized theory can be reformulated as follows.

The anomaly field dataset is a scalar function of $x$ and $z$, where the $x$-axis is the horizontal straight-line at sea level lying in the vertical plane through the survey profile and the $z$-axis is the vertical line, positive downwards. The origin is fixed at sea level half-way through the survey profile (see figure 3). The datum-domain is a surface $S$ (the pseudosection), which can be divided in three sub-domains, $S_1$, $S_2$, $S_3$, as in figure 3, where top and bottom line topography functions are indicated by $z_{i,t}(x)$ and $z_{i,b}(x)$, $i=1,2,3$, respectively. Referring to the definitions of the SPOP and SDOP functions, $\eta_m^{(P)}$ and $\eta_{nv}^{(D)}$, respectively, reported in eq.5a and eq.16a in MP-I, the anomaly field function $A(x,z)$ is defined as

$$A(\mathbf{r}) = \Delta\rho_a(\mathbf{r}) \cong \sum_{m=1}^{M} \frac{\partial \rho_{a,0}(\mathbf{r},\mathbf{r}_m)}{\partial \rho_{m,0}} \Delta\rho_m , \quad (1)$$

where, specifically, $\Delta\rho_a(x,z)$ represents the departure of the measured GAR function $\rho_a(x,z)$ from the synthetic GAR function $\rho_{a,0}(x,z)$ of a reference model, and $\Delta\rho_m$ is the difference between the true resistivity $\rho_m$ in the $m$-th source cell (pole) and the resistivity $\rho_{m,0}$ in the same cell imposed by the reference model. The kernel $s(x,z)$, which takes the role of scanner function, is represented in eq.1 by the Frechet derivative of the synthetic GAR function for a perturbation of the resistivity in the $m$-th cell.

Referring to the geometry sketched in figure 3, the SPOP function used for the tomographies in figure 4 is

$$\eta_m^{(P)} = C_m \int_{S_1+S_2+S_3} f_1 f_2 \, dx \, dz , \quad (2)$$

where

$$C_m = \left( \int_{S_1+S_2+S_3} f_1^2 \, dx \, dz \cdot \int_{S_1+S_2+S_3} f_2^2 \, dx \, dz \right)^{-\frac{1}{2}} \quad (3)$$

In eq.2 and eq.3, $f_1 = \Delta\rho_a(\mathbf{r})$ and $f_2 = \partial \rho_{a,0}(\mathbf{r},\mathbf{r}_m)/\partial \rho_{m,0}$, and the integration intervals along the $x$- and $z$-axis are $[-X,-X_1]$ and $[z_{1,t}(x),z_{1,b}(x)=(Z+X_1)+x]$ in $S_1$, $[-X_1,X_1]$ and $[z_{2,t}(x),Z]$ in $S_2$, $[X_1,X]$, and $[z_{3,t}(x),z_{3,b}(x)=(Z+X_1)-x]$ in $S_3$.

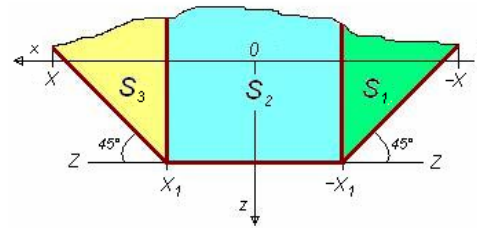

**Figure 3** The *S*-domain partition for the computation of the SPOP and SDOP tomography of the GAR pseudo-sections reported in figure 2.

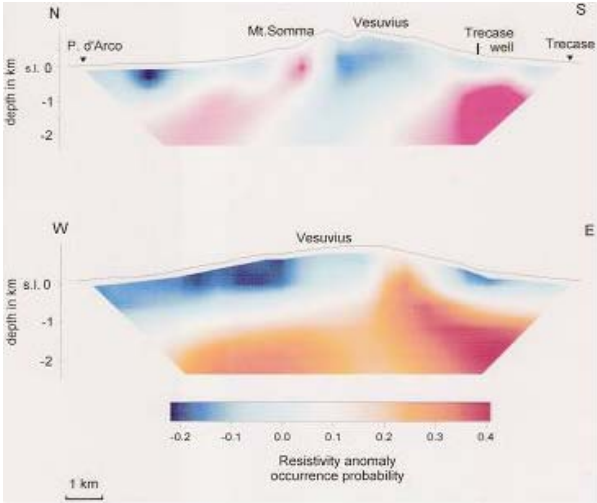

**Figure 4** Geoelectric SPOP tomography imaging at Vesuvius on cross-sections through the N-S (top) and W-E (bottom) profiles shown in figure 1.

The SPOP tomography across the geoelectrical S-N profile shows a very well resolved resistivity anomaly source pattern. Following Patella and Mauriello (1999), the northern shallow negative SPOP nucleus identifies the core of a conductive structure very likely ascribable to marine water-bearing volcano-clastic sediments. The following shallow central positive nucleus, appearing beneath the Mt. Somma caldera northern rim, identifies the core of a resistive structure very likely associable to a slowly cooled compact magmatic dike. The following shallow negative nucleus, located beneath the summit part of the Vesuvius central edifice, identifies the core of a conductive structure, which leads to admit that the volcano chimney is likely occluded by water-rich and mineral-rich fine deposits. Finally, the southern deeper positive nucleus identifies a resistive structure likely associable to the thick sequence of submarine lavas and the underlying carbonatic basement, in agreement with the stratigraphy retrieved from the Trecase well, drilled by the Italian Oil Agency AGIP.

In figure 4, the westward shallow negative nucleus identifies the core of a conductive structure very likely ascribable to a thick aquifer trapped along the western slopes of the volcano. The weak positive nucleus that appears right beneath the Somma caldera eastern rim identifies the core of a resistive body ascribable to the dike previously outlined. No evidence of the central obstructed chimney can be deducted from this profile, probably because the profile, placed to the south of the top cone of Vesuvius, does not cross the bowl-shaped top terminal part of the volcano central conduit, which appears to be slightly displaced northwards. A further shallow negative nucleus appears on the eastern portion of the section. It identifies the core of a conductive layer very likely ascribable to the aquifer trapped along the eastern slopes of the volcano. Finally, the eastern and central-western deeper positive nucleuses highlight the cores of resistive blocks likely referable to east to the faulted carbonate slab outcropping a few km inland, and in the central part to a pile of compact lava flows overlying the said slab, gently downstepping towards the Tyrrhenian sea.

We show now the results of the SDOP tomography. The expressions of $\eta_{n\nu}^{(D)}$ and $C_{n\nu}$ for this geoelectrical survey are written following the procedure used before to derive eq.2 and eq.3. They are for $\nu = x,z$

$$\eta_{n\nu}^{(D)} = C_{n\nu} \int_{S_1+S_2+S_3} f_1 f_3 \, dxdz \,, \qquad (4)$$

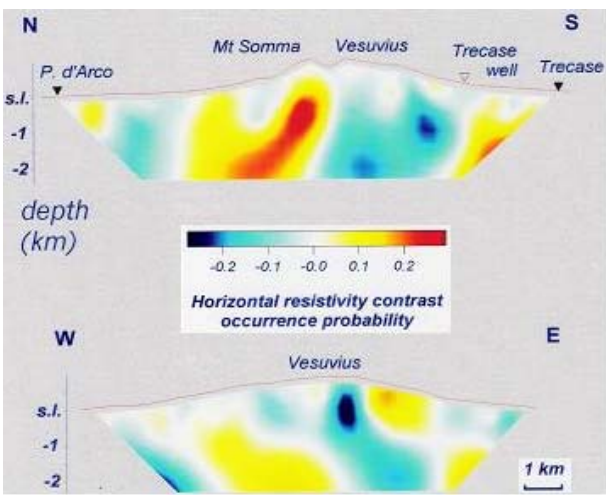

**Figure 5** Geoelectrical *x*-component SDOP tomography at Vesuvius on cross-sections through the S-N (up) and E-W (down) profiles shown in figure 1.

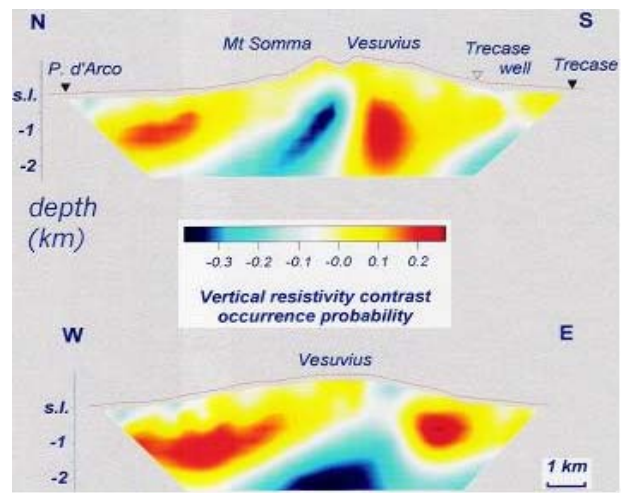

**Figure 6** Geoelectrical *z*-component SDOP tomography at Vesuvius on cross-sections through the S-N (up) and E-W (down) profiles shown in figure 1.

$$C_{n\nu} = \left( \int_{S_1+S_2+S_3} f_1^2 \, dx dz \cdot \int_{S_1+S_2+S_3} f_3^2 \, dx dz \right)^{-1/2}, \qquad (5)$$

where $f_3 = \partial[\partial\rho_{a,0}(\mathbf{r},\mathbf{r}_n)/\partial\rho_{n,0}]/\partial\nu_n$.

Only the two derivatives with respect to $x$ and $z$ can lead to meaningful results, therefore figure 5 and figure 6 show the $\eta_{nx}^{(D)}$ and $\eta_{nz}^{(D)}$ tomographies, respectively.

The $\eta_{nx}^{(D)}$ tomography across the S-N geoelectrical profile of figure 5 depicts a resistivity contrast pattern quite compatible with the previous source pole pattern. The first notable nucleus is that appearing beneath the Mt. Somma – Vesuvius edifice, located exactly amidst the adjacent positive-negative shallow SPOP nucleuses previously discussed. This nucleus, combined with the similar nucleus appearing exactly at the same position in the $\eta_{nz}^{(D)}$ tomography of figure 6, very likely marks the existence of a sharp boundary separating the two different volcanic structures associated to the adjacent SPOP nucleuses. The opposite sign of these two SDOP features and their apparent prolate shape allow the trace of the hypothesized sharp boundary to be imaged in the section nearly as a leftward downverging segment. The second remarkable feature of the $\eta_{nx}^{(D)}$ tomography of figure 5 across the S-N geoelectric profile is the pair of negative cores placed southward between the Vesuvius summit cone and the Trecase well. This pair, together with the nucleus appearing in the $\eta_{nz}^{(D)}$ tomography of figure 6, in the same sector of the section, very likely demarcate the stepped boundary separating the shallow structure of the volcano from the underlying sequence of submarine lavas and carbonates. The sign of the $x$- and $z$-component of these source dipoles is congruent with the sign of the source poles located in that portion of the section. The last remarkable feature of the S-N profile is the positive nucleus perched in the northernmost sector of the $\eta_{nz}^{(D)}$ tomography of figure 6, in the P. d'Arco area. It very likely demarcates the horizontal boundary dividing the shallow water-bearing volcanic sediments from the underlying sequences of compact volcanics.

The E-W $\eta_{nx}^{(D)}$ tomography in figure 5 shows only a negative, vertically prolate nucleus beneath Vesuvius. It very likely highlights a sharp lateral passage from the conductive top central part of Vesuvius to the resistive, nearly vertical wall beneath the eastern rim of the Mt. Somma caldera, interpreted as a compact magma dyke. The E-W $\eta_{nz}^{(D)}$ tomography in figure 6 provides more information. In fact, the western widely prolate positive nucleus may very likely be interpreted as the core of a nearly horizontal boundary dividing the shallow water-bearing volcanic sediments from the underlying pile of compact lava flows. Similarly, the eastern positive core can very likely be interpreted as the trace of a nearly horizontal sharp transition from the superficial water-bearing volcanic sequence to the underlying carbonatic platform. Finally, the deep negative nucleus in the E-W tomography of figure 6 can be associated to the core of the horizontal transition from a pile of compact lavas to a fractured and altered block of the carbonatic slab.

## SELF-POTENTIAL TOMOGRAPHIES

A set of 1250 self-potential (SP) data was collected by the gradient technique displacing a 100 m long passive bipole continuously along a mesh of circuits covering an area of $12 \times 12$ km$^2$ (see figure 1). In figure 7, the top horizontal slice shows the original SP map in mV.

In the case of SP fields, referring to definition given by eq.1 in MP-I, the $\mathbf{A}(\mathbf{r})$ vector function is represented by the steady natural electric field vector on the ground and the SPES vector function $\mathbf{s}(\mathbf{r},\mathbf{r}_m)$ is explicated as

$$\mathbf{s}(\mathbf{r},\mathbf{r}_m) = \frac{\mathbf{r}-\mathbf{r}_m}{|\mathbf{r}-\mathbf{r}_m|^3}. \qquad (6)$$

The SPOP function $\eta_m^{(P)}$ has been calculated using eq.5 and eq.6 in MP-I, and the SDOP function $\eta_{n\nu}^{(D)}$ by eq.16 and eq.17 in MP-I, applying the topography surface regularization function given by eq.12 in MP-I.

In figure 7, the left-hand panel displays the SPOP tomography. In the top slice, a negative closed nucleus appears right in correspondence of the Vesuvius upper cone, circled by positive cores extending downslope to 0,4 km a.s.l., which appear to closely delineate the Mt. Somma caldera rim, mostly its northeastern and southwestern arcs. At sea level, a positive nucleus appears again in correspondence of the Vesuvius cone, together with sequences of aligned positive nucleuses indicating the main radial fractures system crossing the volcanic area. Finally, a roughly circular distribution of positive nucleuses, at sea level, appears to highlight the outer boundary of the volcanic area. This positive pattern at sea level is accompanied by weak negative cores that occupy the intermediate positions and reach the highest values in the slice at 0.4 km of depth b.s.l., first of all that in the northwestern sector.

As is well documented, in hydrothermal systems SP signals are caused by electrokinetic and thermoelectric effects. Generally, in volcanic areas, positive anomalies correspond to upward migrating fluids, while negative anomalies to downward fluid movement (Zlotnicki and Nishida, 2003). Accepting such a model, the top central negative nucleus would outline the position of the main path for meteoric waters infiltration, in correspondence with the Vesuvius crater. The inner crown of positive cores would instead indicate the sites of thermal fluids uprise. This feature might thus represent the signature of a fluid convection mechanism within a thermal field localized above sea level inside the mountain.

The outer crown of positive cores together with the positive radial trends at sea level, may, as previously,

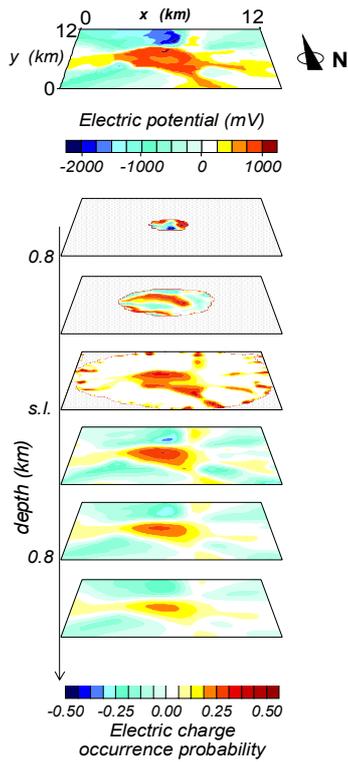

**Figure 7** The SP SPOP tomography of Mt. Vesuvius. The top slice is the reference SP anomaly map.

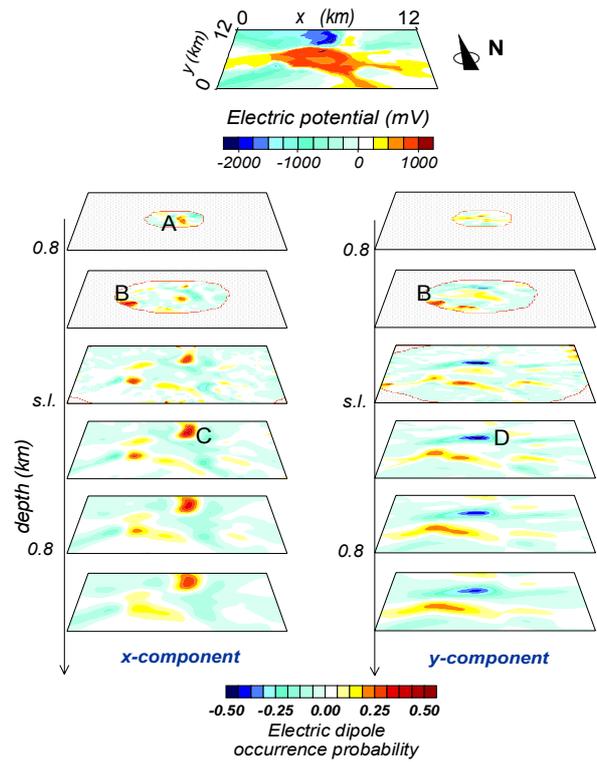

**Figure 8** The SP SDOP tomography of Mt. Vesuvius. The top slice is the reference SP anomaly map.

be assigned to dominant paths for uprising thermal fluids. Accordingly, the negative nucleuses wedged at 400 m of depth b.s.l. inside the positive radial strips are likely to indicate the cores of deep alimentation zones within a larger thermal convection system extended below the whole volcanic area.

The SDOP tomography has been applied to the SP map in order to get a better resolution of the boundary zones. Figure 8 illustrates the SDOP $\eta_{nx}^{(D)}$ and $\eta_{ny}^{(D)}$ tomographies, as only the two derivatives with respect to *x* and *y* can bring to fruitful results. The most evident nucleuses are indicated by a capital letter, which when occurring in both piles indicates that the corresponding nucleuses belong to the same source dipole.

The source "A", appearing on the slice of $\eta_{nx}^{(D)}$ at 0.8 km a.s.l., and the source pair "B", appearing on the slices of $\eta_{nx}^{(D)}$ and $\eta_{ny}^{(D)}$ at 0.4 km a.s.l., would likely indicate the cores of an ideal lateral boundary, dividing the charging area within the Vesuvius conduit from the discharging areas along the northeastern and south-western arcs of the Mt. Somma caldera rim, within the upper thermal system. The sources "C" in the $\eta_{nx}^{(D)}$ tomography and "D" in the $\eta_{ny}^{(D)}$ tomography, reaching the maximum SDOP values at about 0.4 km b.s.l., well outline the boundaries of the wedged feature evidenced in the SPOP tomography.

**GRAVITY TOMOGRAPHIES**

The Bouguer residual gravity map within the Vesuvius area, drawn in the top slice of figure 9, was elaborated by Cassano and La Torre (1987) by means of 1850 data distributed with an average density of 1 station per km$^2$ within an area of 26.6×37 km$^2$. A density of 2.3 g/cm$^3$ was used for slab and terrain corrections and a S35°W trend of 0.9 mgal/km was subtracted from the original values. The main large-scale features are a wide gravity low, extending from the northwestern side to the whole central area, and a gravity high bordering the negative anomaly along the northeastern and southern sectors of the map. The gravity low represents quite well the huge structural depression within which the whole volcanic activity grew, while the gravity high closely depicts the structural highs consisting of the carbonate massifs that border the Campanian plain and reach southwestwards the whole Sorrento peninsula (Principe *et al.*, 1987).

The sequences of slices in figure 9 display the 3D tomography of the SPOP function, where the source poles are assimilated to the centers of mass of bodies with either a surplus or a deficit of mass (Δ-mass) with respect to the hosting medium. In the gravity method, *A*(**r**) represents the scalar Bouguer anomaly dataset (*i.e.* the *z*-component of the gravity field) and the scalar

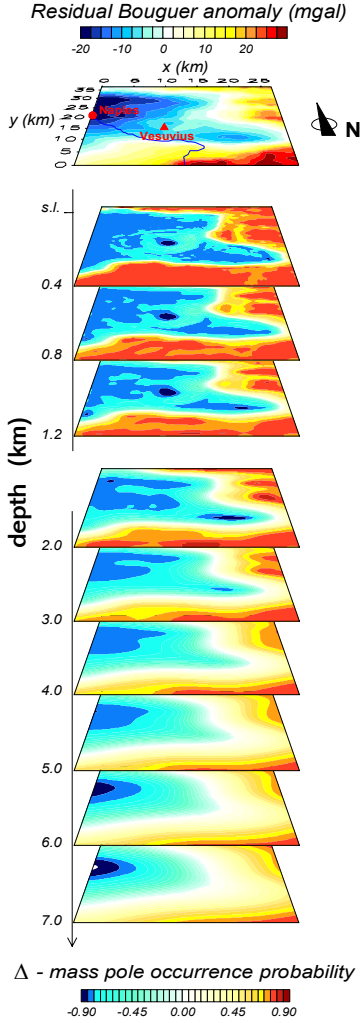

**Figure 9** The gravity SPOP tomography of Vesuvius. The top slice is the reference gravity survey map.

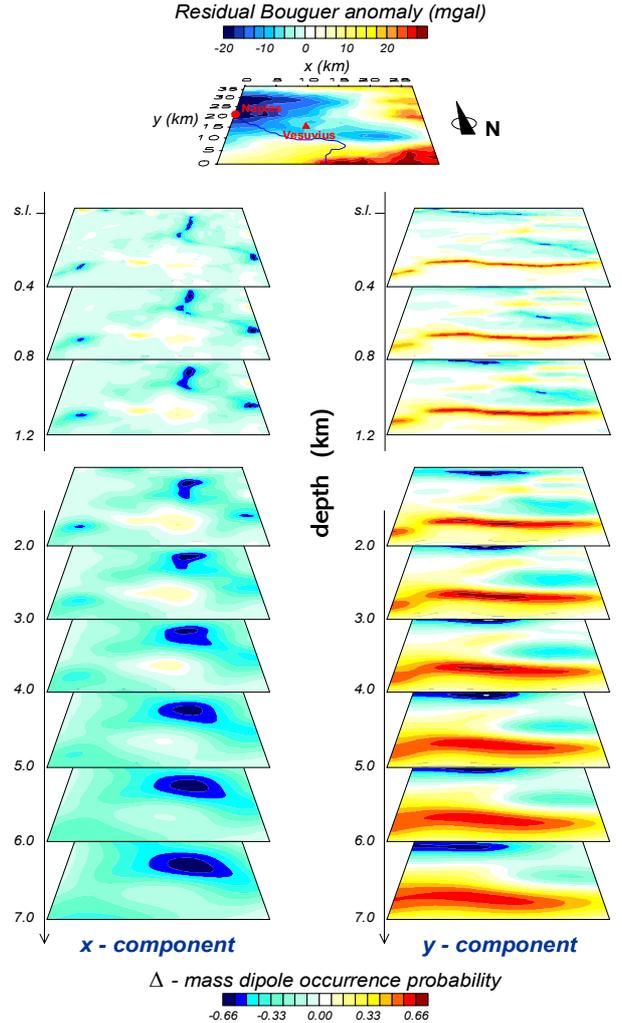

**Figure 10** The gravity SDOP tomography of Vesuvius. The top slice is the reference gravity anomaly map.

SPES function is explicated as

$$s(\mathbf{r},\mathbf{r}_m) = \frac{z_m - z}{|\mathbf{r}_m - \mathbf{r}|^3}. \qquad (7)$$

The SPOP function has been calculated using eq.5a and eq.6 in MP-I, and the SDOP function using eq.16a and eq.17a in MP-I, applying again the topography surface regularization function given by eq.12 in MP-I.

A negative $\Delta$-mass accumulation characterizes the northwestern and middle-eastern part of the surveyed area. Three distinct negative nucleuses aligned SE-NW at different depths emerge inside this huge volume. The easternmost nucleus expands within a depth range of 1.2-2 km b.s.l., while the signature of the northwestern nucleus starts at about 6 km of depth b.s.l. and deepens to more than 7 km b.s.l.. The third core outcrops in the central sector below Vesuvius from about 0.4 km a.s.l. to about 1.6 km b.s.l.. The whole huge depression with the three lighter blocks is surrounded to east and south by a belt of materials denser than 2.3 g/cm$^3$ but with less defined nucleuses. Two denser blocks appear at the eastern and southern margins of the tomospace. Both nucleuses have a very limited extent from 1.6 km b.s.l. down to 2 km b.s.l..

In order to obtain more information, we show now the results of the SDOP tomography used to detect the presence of $\Delta$-mass dipoles ascribable to discontinuity surfaces. Figure 10 shows the SDOP $\eta_{nx}^{(D)}$ and $\eta_{ny}^{(D)}$ tomographies, as only the two derivatives with respect to $x$ and $y$ can bring to meaningful results.

The SDOP tomography provides a detailed image of the boundaries of the huge volume with less dense materials appearing in the northwestern and central-eastern parts of the survey area. The northern, eastern and southern lateral boundaries of this volume are well outlined down to at least 3-4 km b.s.l.. The sign of the SDOP nucleuses reflects the sign of the dipole moment components, conformably to the coordinate reference plane (*x*- and *y*-axis positive eastwards and northwards, respectively).

## CONCLUSION

An application of the probability tomography method to the Vesuvius volcano (Naples, Italy), to image polar and dipolar sources of the geoelectrical, self-potential and gravity anomaly datasets collected over the whole volcanic area, has been shown.

The most important results that have been achieved can be synthesized as follows:
1. a slowly cooled compact magmatic dike is likely to exist beneath the Mt. Somma caldera northern rim;
2. the Vesuvius conduit is likely to be occluded by water-rich and mineral-rich fine deposits;
3. a thick water-bearing horizon is likely to be trapped along the western and eastern slopes of the volcano;
4. a fluid convection mechanism is likely to be active within a thermal field located above sea level inside the Somma-Vesuvius mountain;
5. a larger thermal convection system extended below the whole volcanic area is likely to exist in the first 0.5 km b.s.l.;
6. the whole Mt. Somma-Vesuvius volcanic apparatus is likely to be emplaced in a very large and thick tectonic depression.

We remark that, generally, the areas where the $\eta_m^{(P)}$ and $\eta_n^{(D)}$ functions reach the highest absolute values are, from a probabilistic point of view, the candidate places where to hypothesize the presence of the core-and-boundary sources of the anomalies observed in the datum space. In the Somma-Vesuvius area, the SPOP and SDOP nucleuses with the highest absolute values appear, in fact, as the most suitable places where to ascribe the features listed above. We do not exclude, however, that other points in the tomospace, even if characterized by lower absolute values of the $\eta_m^{(P)}$ and $\eta_n^{(D)}$ functions, if properly combined can represent a potential source model compatible with the observed anomaly field and acceptable from the volcanological point of view.

**Acknowledgements** Study performed with financial grants from the Italian Ministry of Education, University and Research (PRIN 2000 project), the European Commission (TOMAVE project) and the Italian Group of Volcanology of the National Research Council.

## REFERENCES

Cassano E. and La Torre P., 1987. Geophysics. In: R. Santacroce (ed.), Somma-Vesuvius. Quaderni de «La Ricerca Scientifica», 114/8, CNR, Rome, 175-196.

Di Maio R., Mauriello P., Patella D., Petrillo Z., Piscitelli S. and Siniscalchi A., 1998. Electric and electromagnetic outline of the Mount Somma-Vesuvius structural setting. *Journal of Volcanology and Geothermal Research*, **82**, 219-238.

Iuliano T., Mauriello P. and Patella D., 2002. Looking inside Mount Vesuvius by potential fields integrated geophysical tomographies. *Journal of Volcanology and Geothermal Research*, **113**, 363-378.

Mauriello P. and Patella D., 1999. Resistivity anomaly imaging by probability tomography. *Geophysical Prospecting*, **47**, 411-429.

Mauriello P. and Patella D., 2006. Imaging polar and dipolar sources of geophysical anomalies by probability tomography. Part I: theory and synthetic examples. http://arxiv.org/physics/0602056, 1-6.

Patella D., 1997. Self-potential global tomography including topographic effects. *Geophysical Prospecting* **45**, 843-863.

Patella D. and Mauriello P., 1999. The geophysical contribution to the safeguard of historical sites in active volcanic areas. The Vesuvius case-history. *Journal of Applied Geophysics*, **41**, 241-258.

Principe C., Rosi M., Santacroce R. and Sbrana A., 1987. Explanatory notes to the geological map. In: R. Santacroce (ed.), Somma-Vesuvius. Quaderni de «La Ricerca Scientifica», 114/8, CNR, Rome, 11-51.

Spera F.J., De Vivo B., Ayuso R.A. and Belkin H.E. (ed.s), 1998. Special issue: Vesuvius. *Journal of Volcanology and Geothermal Research*, **82**, 247 p.

Zlotnicki J. and Nishida Y., 2003. Review on morphological insights of self-potential anomalies on volcanoes. *Surveys in Geophysics*, **24**, 291-338.